\documentclass[12pt,preprint]{aastex}

\newcommand{\myemail}{nasghar@dubs.ac.ir}

\slugcomment{This paper has been accepted by \textbf{MNRAS}}

\shorttitle{Isobaric Thermal Instability}
\shortauthors{Nejad-Asghar}

\begin{document}


\title{Formation of Fluctuations in the Molecular Slab via Isobaric Thermal Instability}


\author{Mohsen Nejad-Asghar\altaffilmark{1,2}}

\affil{$^1$Department of Physics, Damghan University of Basic
Sciences, Damghan, Iran}

\affil{$^2$Research Institute for Astronomy and Astrophysics of
Maragha, Maragha, Iran}

\email{\myemail}


\begin{abstract}
The frictional heating by ion-neutral drift is calculated and its
effect on isobaric thermal instability is carried out. Ambipolar
drift heating of one-dimensional self-gravitating magnetized
molecular slab is used under the assumptions of
quasi-magnetohydrostatic and local ionization equilibrium. We see
that ambipolar drift heating is inversely proportional to density
and its value in some regions of the slab can be significantly
larger than the average heating rates of cosmic rays and
turbulent motions. The results show that the isobaric thermal
instability can occur in some regions of the slab; therefore it
may produce the slab fragmentation and formation of the AU-scale
condensations.
\end{abstract}





\section{Introduction}
Contour maps of the molecular clouds show a very fragmented
distribution, consisting several clumps and cores (see, e.g.,
Williams et al. 2000, Falgarone et al. 2004). Cloud structures are
likely to be more complex than this because many of the clumps
that are optically thick in one transition contain substructure
that is visible in other transitions (Perault et al. 1985, Kwan \&
Sanders 1986, Dickman et al. 1990).

The reality of AU-scale structure within molecular gas is still
questionable. Indeed, only minor species can be easily observed
towards molecular clouds and their spatial distribution might not
reflect that of $\mathrm{H_2}$ were present. $\mathrm{H_2CO}$,
$\mathrm{HCO^+}$, and $\mathrm{OH}$ apparently display column
density fluctuations reaching $5$ to $15\%$ along lines of sight
separated by $10$~AU (Marscher et al. 1993, Moore \& Marscher
1995, Liszt \& Lucas 2000), while dust grains appear to be more
smoothly distributed (Thoraval et al. 1996). Rollinde et al.
(2003) probed the spatial distribution of $\mathrm{CH}$,
$\mathrm{CH^+}$, and $\mathrm{CN}$ at scales ranging from $1$~AU
up to $150$~Au. Studies of the time variability of absorption
lines indicates the presence of fluctuations on scales of $10^{-4}
\mathrm{pc}$ (5-50 AU) and masses of $10^{-9} \mathrm{M_\odot}$
(Boiss\'{e} et al. 2005). At larger scales (about $10,000$~AU),
Pan et al. (2001) find significant differences in $\mathrm{CN}$,
$\mathrm{CH}$, and $\mathrm{CH^+}$ absorption lines.

Star forming regions are often interpreted in a picture that
consists of an ensemble of discrete regions of gas. Existence of
these AU-scale condensations in the clumps of the molecular
clouds, may be a motivation to investigate the hierarchy down-up
structure of the clouds. Various authors, such as Kwan~(1997) and
Silk \& Takahashi~(1997), have cited coagulation processes as an
origin for the form of the initial mass function (IMF). For
example, Murray \& Lin~(1996) have argued that in protoglobular
clusters, gas is fragmented onto sub-Jeans-mass clumps by a
variety of hydrodynamical instabilities, and that eventual star
formation ensues when enough clumps have merged to form Jeans
unstable objects. Pringle et al.~(2001) have considered a similar
process in the assembly of giant molecular clouds in the spiral
arms of disc galaxies. Thus, the condensations appear, as a
possible scenario, to be precursors of large-scale cores (dense
cores with significant jeans mass) via merging and collisions;
they constitute the initial conditions for star formation. This
idea is largely an attempt to characterize the observed structure
of molecular cloud complexes- see, e.g. the review by Williams et
al.~(2000)- but is likely to be applicable in many situations.

It is well known that molecular clouds are turbulent which
triggers the formation of structures. A recent literature has
suggested that the clumpiness in clouds arises naturally from
their formation through thermal instability, which acts on
time-scales that can be much shorter than the duration of the
turbulent motions. This suggests that molecular clouds may be
already fragmented when they form (Koyama \& Inutsuka~2002, Audit
\& Hennebelle~2005, Heitsch et al.~2006, Vazquez-Semadeni et
al.~2006). In the same way, other mechanisms to generate
substructures inside molecular clouds, based on MHD waves, have
been proposed (Gammie \& Ostriker~1996, Falle \& Hartquist~2002,
Folini et al.~2004). In this paper we make the assumption that the
molecular cloud is initially an uniform ensemble which then
fragments due to thermal instability. This is clearly a
competitive mechanism that we considered although all could
operate.

It is also believed that the interstellar medium is strongly
magnetized. Although the structure of the magnetic field is
poorly known, various studies have considered thermal instability
and magnetic field as well. Loewenstein~(1990) investigated the
effect of magnetic fields on thermal instability in cooling flows
using linear perturbation analysis. Steele \&
Ib\'{a}\~{n}ez~(1999) consider the nonlinear thermal disturbances
for a two-dimensional structure taking into account thermal
conduction parallel to and perpendicular to the magnetic field.
Hennebelle \& Per\'{a}ult~(2000) have investigated the role of an
initially uniform magnetic field, analytically and numerically,
when the thermal conduction is dynamically triggered. They
propose a mechanism based on magnetic tension to explain the
thermal collapse in a magnetized flow and argue that the magnetic
intensity in the WNM and in the CNM should not be very different.
The effect of Alfv\'{e}n waves on thermal instability was studied
by Hennebelle \& Passot~(2006). Inoue et al.~(2007) Performed
one-dimensional two-fluid numerical simulations that describe the
neutral and ionized components in the interstellar medium with
purely transverse magnetic fields. They find that the cloud that
is formed by the thermal instability always has magnetic field
strength on the order of a few $\mu$G irrespective of the initial
strength.

In molecular clouds, a neutral molecular gas intermixed with an
ionized component that is tied directly to the magnetic field.
Birk~(2000) has used the two-fluid technique to find the thermal
condensation modes in weakly ionized hydrogen plasma. Nejad-Asghar
\& Ghanbari~(2003) studied the effect of linear thermal
instability in a weakly ionized magnetic molecular cloud within
the one-fluid description. They concluded that not only the
thermal instability is an important mechanism but also there are
solutions where the isentropic thermal instability including
ambipolar diffusion allows compression along the magnetic field
but not perpendicular to it. Stiele et al.~(2006 hearafter SLH)
revisit carefully the above scenario with some improvements. They
drive a critical wavelength that separates the ranges of
stability and instability. Moreover, they discussed on the
relations between the parameters of cooling and heating rates,
where isobaric and/or isentropic thermal instability can be
occurred. Nejad-Asghar \& Ghanbari~(2006) investigated the effect
of self-gravity in occurrence of the thermal instability in a
weakly ionized magnetic molecular cloud. They concluded that
there are, not only oblate condensation forming solutions, but
also prolate solutions according to expansion or contraction of
the clump.

The goal of this paper is to analyze the effect of ambipolar
drift heating on net cooling function and investigation of the
isobaric thermal instability in a self-gravitating magnetized
molecular slab. We re-formulate the equations of a
quasi-magnetohydrostatic self-gravitation slab in \S~2. Then,
density dependence of the drift velocity is investigated.
Section~3 deals with the net cooling rate of a standard molecular
cloud as discussed by Goldsmith~(2001, hereafter G01). Moreover,
we investigate occurrence of the isobaric thermal instability and
formation of the AU-scale condensations in some regions of the
molecular slab. Finally, \S~4 allocates to a gathering summary
with conclusion.

\section{Self-Gravitating Slab}
\begin{figure}
\epsscale{.50} \plotone{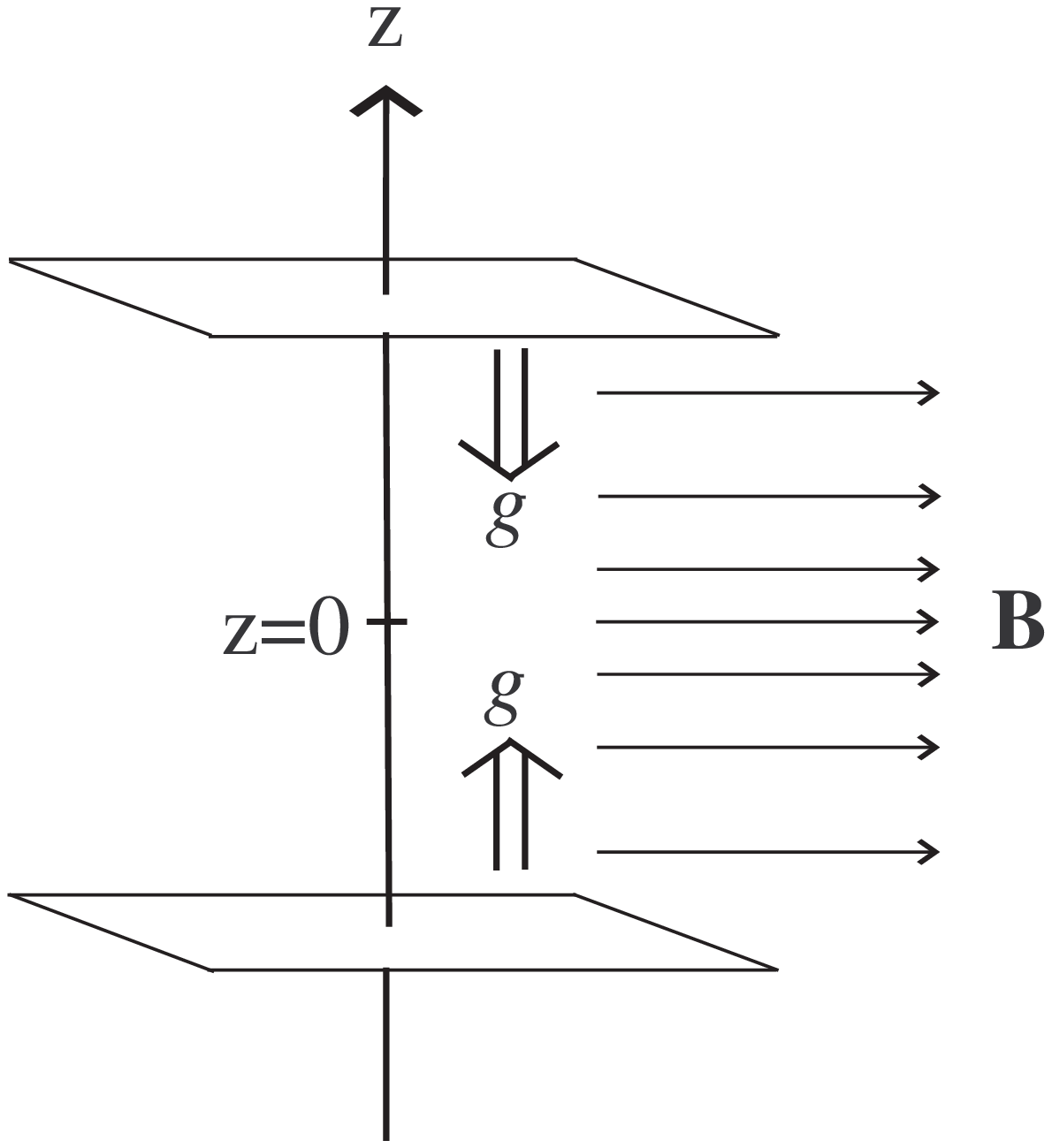} \caption{Schematic diagram of
self-gravitating magnetic slab in $z$ direction. \label{fig1}}
\end{figure}

We consider the evolution of a plane-parallel self-gravitating
magnetized slab of lightly ionized molecular gas (see,
Fig.~\ref{fig1}). In this geometry, all variables are functions of
distance $z$ to the central plane and time $t$ only. The magnetic
field is frozen only to the ions, so that diffusion of the field
relative to the neutral gas must continuously insert a drag force
(per unit volume)
\begin{equation}\label{dforce}
  f_d = \gamma_{AD} \rho_i \rho v_d,
\end{equation}
where $\gamma_{AD} \sim 3.5 \times 10^{10} \mathrm{m^3/kg.s}$
represents the collision drag coefficient and $\rho_i = \epsilon
\rho^\nu$ is the ion density. In ionization equilibrium state
$\nu=1/2$ and $\epsilon\sim 9.5\times 10^{-15}
\mathrm{kg^{1/2}.m^{-3/2}}$ is valid, and $\nu\neq 1/2$ represents
a deviation from equilibrium state (Shu~1992). The drift velocity
is given by
\begin{equation}\label{drift}
  v_d = -\frac{1}{\gamma_{AD} \epsilon \rho^{1+\nu}}\frac{\partial}{\partial
  z} (\frac{B^2}{2\mu_0}),
\end{equation}
which is obtained by assumption that the pressure and
gravitational force on the charged fluid component are negligible
compared to the Lorentz force because of the low ionization
fraction. In this way, the magnetic fields are directly evolved
by charged fluid component, as follows:
\begin{equation}\label{magcon}
  \frac{d B}{d t} = - B \frac{\partial v}{\partial z} +
  \frac{\partial}{\partial z} (B v_d).
\end{equation}

Since the ion density is negligible in comparison to the neutral
density, the mass conservation is given by
\begin{equation}\label{mascon}
\frac{d \rho}{d t} =- \rho \frac{\partial v}{\partial z}.
\end{equation}
The momentum equation then becomes
\begin{equation}\label{momcon}
  \frac{d v}{d t} = g - \frac{1}{\rho} \frac{\partial}{\partial z} (a^2 \rho +
  \frac{B^2}{2\mu_0})
\end{equation}
where $a$ is the isothermal sound speed and the gravitational
acceleration $g$ obeys the poisson's equation
\begin{equation}\label{poisson}
  \frac{\partial g}{\partial z} = -4 \pi G \rho.
\end{equation}

Following the many previous treatments, a further simplification
is possible if we introduce the surface density between mid-plane
and $z>0$ as
\begin{equation}\label{sig}
\sigma \equiv \int_{0}^{z} \rho(z',t)dz'.
\end{equation}
The usual defined surface density from $-z$ to $+z$ has twice the
value of $\sigma$. By transformation from $(z,t)$ to $(\sigma,
t)$,
\begin{eqnarray}
\nonumber \frac{d}{d t} = \left( \frac{\partial}{\partial t}
\right)_\sigma , \quad \frac{\partial}{\partial z} = \rho \left(
\frac{\partial}{\partial\sigma} \right)_t,
\end{eqnarray}
the drift velocity is
\begin{equation}\label{drift2}
  v_d = -\frac{1}{\gamma_{AD} \epsilon \rho^\nu}\frac{\partial}{\partial
  \sigma} (\frac{B^2}{2\mu_0}),
\end{equation}
and the equation (\ref{magcon}) becomes
\begin{equation}\label{magcon2}
  \frac{\partial}{\partial t} (\frac{B}{\rho})= \frac{1}{\gamma_{AD}\epsilon\mu_0}
  \frac{\partial}{\partial \sigma} (\frac{B^2}{\rho^\nu}\frac{\partial B}{\partial
  \sigma}).
\end{equation}
With the above, field equation (\ref{poisson}) can be integrated
to give
\begin{equation}\label{poisson2}
  g = - 4 \pi G  \sigma,
\end{equation}
while the equation of continuity (\ref{mascon}) and the equation
of motion (\ref{momcon}) take the form
\begin{equation}\label{mascon2}
  \frac{\partial z}{\partial \sigma} = \frac{1}{\rho}
\end{equation}
and
\begin{equation}\label{momcon2}
  \frac{\partial^2 z}{\partial t^2} = -4 \pi G \sigma
   - \frac{\partial}{\partial \sigma} (a^2 \rho+ \frac{B^2}{2\mu_0}),
\end{equation}
respectively. The slab is assumed to be in
quasi-magnetohydrostatic equilibrium at all time, supported
against its own self-gravity by the magnetic nd gas pressure. The
loss of flux from ambipolar diffusion is exactly compensated for
by the compression of the slab which is necessary to maintain
equilibrium. In this approximation, the left-hand side of equation
(\ref{momcon2}) is zero, and we may integrate the force balance
to obtain
\begin{equation}\label{momcon3}
  \frac{B^2}{2\mu_0} + a^2 \rho = 2\pi G (\sigma_\infty^2 - \sigma^2)
\end{equation}
where integration constant $\sigma_\infty$ is the value of
$\sigma$ at $z=\infty$ (where $\rho$ is zero).

In a sense, our goal is to obtain the variation of drift velocity
as a function of density at initial time. Evolution of the slab
is beyond the scope of this paper. For this purpose, we follow
the work of Shu~(1983) to introduce the non-dimension quantities
\begin{eqnarray}\label{nond}
  \nonumber \tilde{\sigma} \equiv \frac{\sigma}{\sigma_\infty} , \quad
  \tilde{\rho} \equiv \frac{a^2}{2 \pi G \sigma_\infty^2} \rho,\quad
  \tilde{z} \equiv \frac{2 \pi G \sigma_\infty}{a^2} z, ~~~~~\\
  \tilde{B} \equiv \frac{B}{2 \sigma_\infty \sqrt{\pi \mu_0 G}} ,\quad
  \tilde{t} \equiv (\frac{2(2\pi G)^{1-\nu}}{\gamma_{AD} \epsilon}) (\frac{2\pi
  G\sigma_\infty^{2-2\nu}}{a^{2-2\nu}})t,
\end{eqnarray}
so that we rewrite the basic equations (\ref{magcon2}),
(\ref{mascon2}) and (\ref{momcon3}) as follows:
\begin{equation}\label{magcon4}
  \frac{\partial}{\partial \tilde{t}}
  (\frac{\tilde{B}}{\tilde{\rho}})= \frac{\partial}{\partial
  \tilde{\sigma}}(\frac{\tilde{B}^2}{\tilde{\rho}^\nu} \frac{\partial
  \tilde{B}}{\partial\tilde{\sigma}}),
\end{equation}
\begin{equation}\label{momcon4}
  \tilde{B}^2 + \tilde{\rho}= 1- \tilde{\sigma}^2,
\end{equation}
\begin{equation}\label{mascon4}
  \frac{\partial \tilde{z}}{\partial
  \tilde{\sigma}}=\frac{1}{\tilde{\rho}}.
\end{equation}
In this way, a natural family of initial states is generated by
assuming that the initial ratio of magnetic to gas pressure is
everywhere a constant, $\alpha_0$, i.e.,
$\tilde{B}^2/\tilde{\rho} = \alpha_0$ at $\tilde{t} =0$.  Then
one finds from equations (\ref{momcon4}), (\ref{mascon4}) and
(\ref{drift2}) that
\begin{equation}\label{dens}
  \rho _{(z,t=0)} = \frac{\rho_0}{\cosh^2(z/z_\infty)},
\end{equation}
\begin{equation}\label{drift3}
  v_d = \frac{2\alpha_0}{\sqrt{1+\alpha_0}} \frac{a \sqrt{2\pi G}}{\gamma_{AD} \epsilon \rho_0^{\nu-1/2}}
  \left(\frac{\rho_{(z,t=0)}}{\rho_0}\right)^{-\nu} \sqrt{1-\frac{\rho_{(z,t=0)}}{\rho_0}},
\end{equation}
where $\rho_0 \equiv 2\pi G \sigma_\infty^2 / a^2 (1+\alpha_0)$ is
the central density of the slab at $t=0$, and $z_\infty \equiv a
\sqrt{(1+\alpha_0)/2\pi G \rho_0}$ is a length-scale parameter.

\begin{figure}
\epsscale{1.0} \plotone{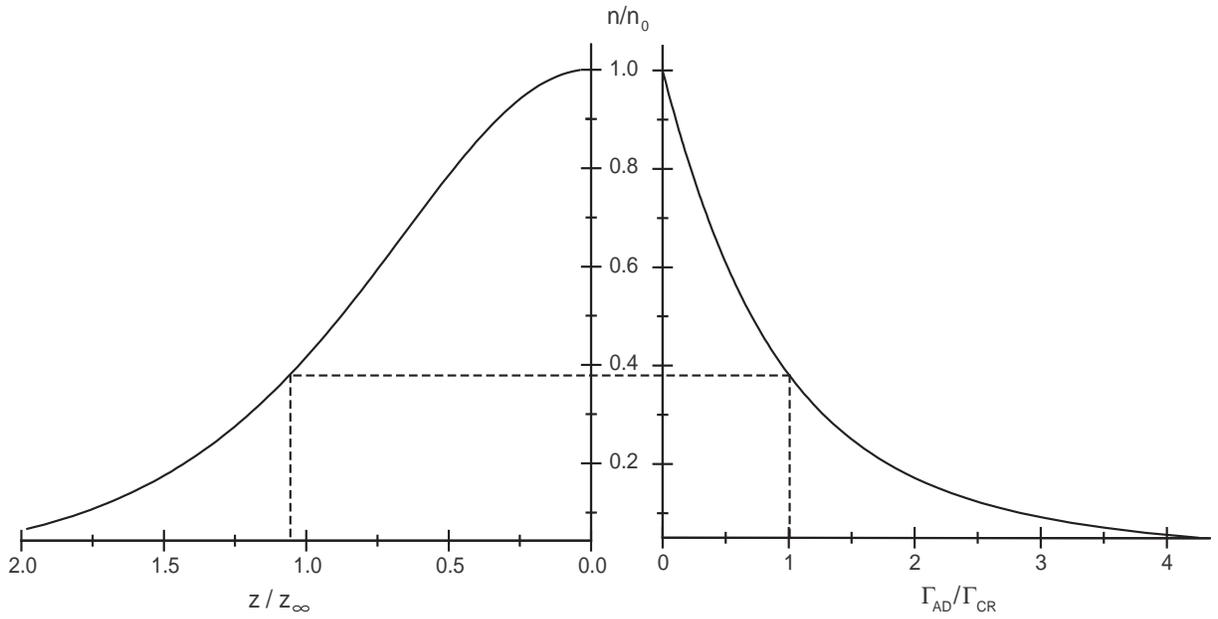} \caption{Relative magnitude of
the ambipolar drift heating rate of a self-gravitating molecular
slab in the case of $\alpha_0=1$, $a=0.3\mathrm{km.s^{-1}}$, and
$n_0=10^{12}\mathrm{m^{-3}}$. The density profile of the slab is
given for comparison. \label{fig2}}
\end{figure}

The interior magnetic field of a molecular cloud acts only on the
ions and electrons, so that the tendency for the field lines to
straighten out is impeded by collisions of charged particles with
the neutral particles. A simple physical derivation of the
ambipolar drift heating rate ($\mathrm{J.kg^{-1}.s^{-1}}$) is
given by considering the drag force (\ref{dforce}) as follows
\begin{equation}\label{heatAD}
  \Gamma_{AD} = \frac{\textbf{f}_d.\textbf{v}_d}{\rho} =
  \gamma_{AD} \epsilon \rho^\nu v_d^2.
\end{equation}
Substituting the equations (\ref{dens}) and (\ref{drift3}) into
equation (\ref{heatAD}), the ambipolar drift heating rate of a
molecular slab is given by
\begin{equation}\label{heatAD2}
  \Gamma_{AD}= \frac{\alpha_0^2}{1+\alpha_0} \frac{8\pi G}{\gamma_{AD} \epsilon}
  a^2 \rho_0^{1-\nu} \frac{1-\rho/\rho_0}{(\rho/\rho_0)^\nu},
\end{equation}
which can be compared to the heating of cosmic rays (e.g.,
Goldsmith~2001),
\begin{equation}\label{heatCR}
  \Gamma_{CR} \approx 3.12 \times 10^{-8}\quad \mathrm{J.kg^{-1}.s^{-1}},
\end{equation}
as follows ($\nu=1/2$)
\begin{equation}\label{ADtoCR}
  \frac{\Gamma_{AD}}{\Gamma_{CR}} = \frac{\alpha_0^2}{1+\alpha_0} \left(\frac{a}{300
  \mathrm{m.s}^{-1}}\right)^2 \left(\frac{n_0}{10^{12} \mathrm{m}^{-3}}\right)^{1/2} \frac{1-n/n_0}{(n/n_0)^{1/2}}
\end{equation}
that is shown in Fig.~\ref{fig2} for typical values of
$\alpha_0=1$, $a=0.3\mathrm{km.s^{-1}}$, and
$n_0=10^{12}\mathrm{m^{-3}}$. We have chosen to plot the density
profile (\ref{dens}), because this makes it easy to compare the
relative magnitude of ambipolar drift heating at different
regions of the slab. Since the drift velocity is inversely
proportional to the density (e.g., Eq.\ref{drift3}), it is clear
that the ambipolar drift heating rate is predominated in the
outer regions of the slab.

It would also be interesting to derive a rough estimate of
heating due to the drift induced by turbulent motions and to
compare its order of magnitude with the heating of cosmic rays.
Following Black~(1987) the turbulence dissipation heating rate
can be estimated as
\begin{eqnarray}\label{turb}
  \nonumber \Gamma_{turb} &\sim& \frac{(\frac{1}{2} m v_{turb}^2)
  (v_{turb}n) (\frac{1}{l})}{mn} \\
  &\sim& 1.6 \times 10^{-8} (\frac{v_{turb}}{1~\mathrm{km.s}^{-1}})^3
  (\frac{1~\mathrm{pc}}{l}) \quad
  \mathrm{J.kg}^{-1}.\mathrm{s}^{-1},
\end{eqnarray}
where $v_{turb}$ is the turbulent velocity and $l$ is the
dissipation length scale. With $v_{turb} \sim 1~
\mathrm{km.s}^{-1}$ and $l\sim 1~\mathrm{pc}$, we calculate
$\Gamma_{turb} \sim 1.6 \times 10^{-8}
\mathrm{J.kg}^{-1}.\mathrm{s}^{-1}$, comparable to half of the
heating rate of cosmic rays (i.e., Eq.~\ref{heatCR}).

\section{Isobaric Thermal Instability}

\subsection{The Molecular Gas Cooling Function}

The cooling rate for molecular gas that is used in this work is
based on the "standard abundances" adopted by G01. I will briefly
review here the important aspects of its construction.
Calculation of thermal balance in molecular clouds by Goldsmith
\& Langer~(1978), Neufeld et~al.~(1995), and G01 have shown that
the various isotopic variants of the carbon mooxide molecule are
certainly important coolants at high densities (i.e. $n\gtrsim
10^{10} \mathrm{m^{-3}}$). This is because the transitions that
contribute to the cooling are thermalized due to the combination
of higher collision rate and radiative trapping. The "standard
abundances" for $^{12}\mathrm{CO}$, $^{13}\mathrm{CO}$, and
$\mathrm{C^{18}O}$ was adopted by G01 based on the observational
data equal to $5.62 \times 10^{-5}$, $1.00 \times 10^{-6}$, and
$1.00 \times 10^{-7}$ relative to that of $\mathrm{H}_2$,
respectively. Neutral Carbon is an important coolant since
although it has only two fine structure transitions available,
they have modest energies, and the abundance of this atom is
relatively large.

The fractional abundances of ortho and para water vapor in
molecular clouds are surprisingly low, so that G01 adopted a
number density of ortho-$\mathrm{H}_2\mathrm{O}$ relative to that
of $\mathrm{H}_2$ equal to $10^{-8}$. He have also doubled the
ortho-$\mathrm{H}_2\mathrm{O}$ cooling rate to account
approximately for contribution of para-$\mathrm{H}_2\mathrm{O}$.
There is a large number of molecular species, including
$\mathrm{CN}$, $\mathrm{HCN}$, $\mathrm{CS}$, $\mathrm{NH}_3$,
$\mathrm{HCO}^+$, and $\mathrm{H}_2 \mathrm{CO}$, that have low
fractional abundances. To simplify the cooling calculation, G01
proposed the $\mathrm{CS}$ molecule as representative and
multiplied its cooling by a factor of $10$ to include the
contribution of the other species in this category.

The velocity field in molecular clouds is not well understood in
detail. We can assume that the large velocity gradient
approximation is valid for the radiative transfer calculations
necessary to compute the molecular cooling when the optical
depths are not small. A critical input to the large velocity
gradient calculation is the molecular abundance per unit velocity
gradient, which is equivalent to the column density per unit line
width. Comparing the sizes and line widths for many clouds gives a
characteristic value of $1~\mathrm{km.s}^{-1}.\mathrm{pc}^{-1}$
that was used by G01. The use of the large velocity gradient
model introduces some uncertainty, but, do not significantly
alter the total cooling rate.

In this way, the total cooling has a complicated dependence on
the temperature and density of molecular hydrogen, parameterized
by G01 as
\begin{equation}\label{cool}
  \Lambda_{(n,T)} = \Lambda_{(n)} (\frac{T}{T_0})^{\beta(n)} \quad
  \mathrm{J. s^{-1}. kg^{-1}},
\end{equation}
where $T_0=10\mathrm{K}$ and the values of $\Lambda(n)$ and
$\beta(n)$ are given in Table~1 for standard abundances and
velocity gradient of $1~ \mathrm{km.s}^{-1}.\mathrm{pc}^{-1}$.
Here, we use the polynomial fitting functions as follows
\begin{equation}\label{lambda0}
  \log\left(\frac{\Lambda_{(n)}}{\mathrm{J.kg^{-1}.s^{-1}}}\right) = -8.98 -
  0.87 (\log \frac{n}{n_0}) - 0.14 (\log \frac{n}{n_0})^2,
\end{equation}
\begin{equation}\label{beta}
  \beta_{(n)} = 3.07 - 0.11 (\log \frac{n}{n_0}) - 0.13 (\log \frac{n}{n_0})^2,
\end{equation}
where $n_0=10^{12} \mathrm{m^{-3}}$. Another expression for the
parameter $\Lambda_{(n)}$ is
\begin{equation}\label{lambda02}
  \log \Lambda_{(n)} = \log \Lambda_0 + \delta \log \frac{n}{n_0},
\end{equation}
which was used by Nejad-Asghar \& Ghanbari~(2003, 2006) and SLH,
for linear investigation of thermal instability in a narrow range
of density. In this formalism, the approximate values of
$\Lambda_0$ and $\delta$ between neighbor densities are given in
Table~1.

There is observational evidences that a variety of molecular
species are depleted on the surface of dust grains (e.g., Tielen
\& Whittet~1997). The observational results on depletion of
different species are relatively limited and are certainly
compromised by blending of emission from different regions along
the line of sight. It is quite difficult to know exactly how the
column densities or abundances of species observed on dust grain
surfaces are related to the abundances in the gas phase. There
also have not been systematic studies of the depletion of a wide
range of molecular species in a set of sources. In this way, G01
used the fragmentary data from a selection of clouds to analyze
the effects of molecular depletion on the standard values of the
cooling function parameters $\Lambda(n)$ and $\beta_{(n)}$.
Because of the above mentioned uncertainties, we should treat the
cooling function in a parametric fashion with $\Lambda(n)$ and
$\beta_{(n)}$ approximately near the standard values of Table~1.

The cooling function was constructed assuming that statistical
steady-state equilibrium is valid, so that the chemical times are
very short with respect to dynamical and cooling times. For all
of the trace coolants this is reasonable except $\mathrm{H}_2$
molecule that the collision time (time to reach statistical
steady-state equilibrium) is necessarily of order the cooling
time (Gilden~1984). The $\mathrm{H}_2$ is the main coolant at
higher temperatures (the temperature where the carbon monoxide
and $\mathrm{H}_2$ contribute equally is slowly increasing
function of density). At the very low temperatures of interest
considered by G01 in the molecular clouds, $\mathrm{H}_2$ do not
play a significant role as a major coolant because of its
low-lying energy levels. Thus, the standard values of the cooling
function parameters $\Lambda(n)$ and $\beta_{(n)}$ (as mentioned
in Table~1) based on the statistical steady-state equilibrium are
approximately appropriate for our analysis.

\begin{figure}
\epsscale{1.0} \plotone{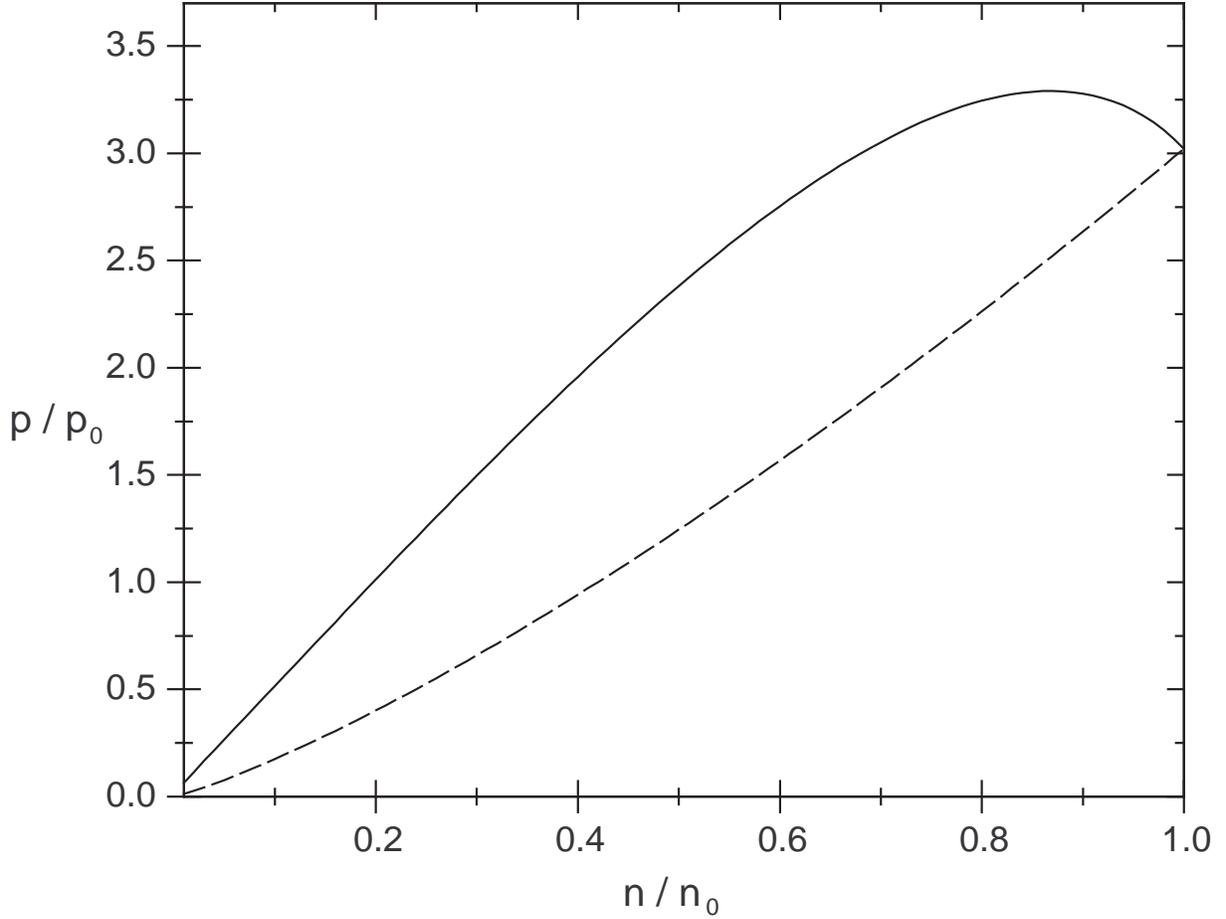} \caption{Curves of thermal
equilibrium in the pressure-density plane. The dash curve
represents the equilibrium state without magnetic field (i.e.,
$\alpha_0=0$), and the solid curve is obtained by considering the
ambipolar drift heating rate with $\alpha_0=10$,
$a=0.3\mathrm{km.s^{-1}}$, and $n_0=10^{12}\mathrm{m^{-3}}$.
\label{fig3}}
\end{figure}

Models of the molecular clouds identify several different heating
mechanisms. Here, we use the heating of cosmic ray (\ref{heatCR})
and the ion-neutral slip heating rate as described by equation
(\ref{ADtoCR}). The isobaric instability mode in a thermal
equilibrium gas can be discussed in relation to the two-phase
model in the same way as described in figure~3 of Gilden~(1984).
The detailed study of the linear investigation and dynamical
evolution of the isobaric modes are given in the next
sub-section. The instance here can be analyzed in a graphical
approach. The pressure-density plane is an elegant method to
describe the isobaric instability modes in the thermal
equilibrium,
\begin{equation}\label{equilib}
  \Lambda_{(n)}(\frac{T}{T_0})^{\beta(n)} = \Gamma_{CR}+\Gamma_{AD},
\end{equation}
of a molecular gas. Substituting the equation (\ref{ADtoCR}) into
(\ref{equilib}), the pressure in thermal equilibrium
self-gravitating molecular slab can be obtained as
\begin{eqnarray}\label{pre}
  \nonumber \frac{p}{p_0}&=&\frac{n T}{n_0 T_0}
  \\
  &=&(\frac{n}{n_0}) \left\{\frac{\Gamma_{CR}}{\Lambda_{(n)}}
  \left[1 + \frac{\alpha_0^2}{1+\alpha_0} (\frac{a}{300
  \mathrm{m.s}^{-1}})^2 (\frac{n_0}{10^{12} \mathrm{m}^{-3}})^{1/2}
  \frac{1-n/n_0}{(n/n_0)^{1/2}} \right] \right\}^{1/\beta(n)}.
\end{eqnarray}
The two-phase instability is displayed in Fig.~\ref{fig3} where
the values of $a$ and $n_0$ are typically chosen equal to
$0.3\mathrm{km.s^{-1}}$ and $10^{12}\mathrm{m^{-3}}$,
respectively. The diagram of Fig.~\ref{fig3} shows the equilibrium
state without magnetic field (i.e., $\alpha_0=0$) in comparison
with considering of the ambipolar drift heating rate with
$\alpha_0=10$. This figure suggests that without ambipolar
diffusion friction, there would not be isobaric thermal
instability. This is not what has been found by Gilden~(1984),
because as we known the thermal instability is very sensitive to
the net cooling function. Here we use the approximately most
accurate cooling function which is evaluated by G0. An
equilibrium state is specified by the intersection of a thermal
equilibrium curve and a constant pressure line. The two curves in
Fig.~\ref{fig3} display that considering of ambipolar drift
heating rate causes to bifurcate the slab into two phases, with
individual gas elements going into that phase corresponding to
the sign of the initial fluctuation.

\subsection{Condensation Growth}
\begin{figure}
\epsscale{1.0} \plotone{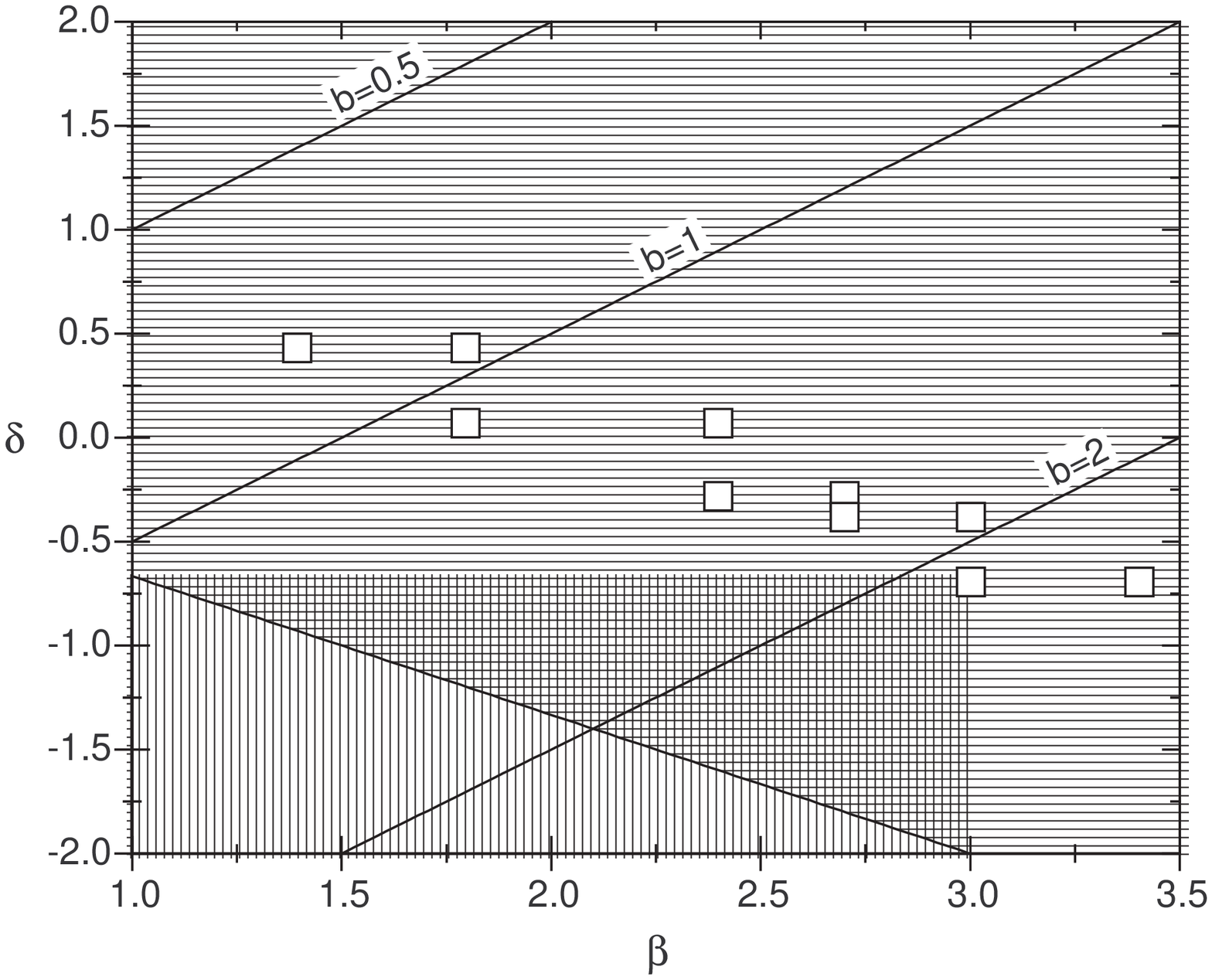} \caption{Isobaric (horizontal
shade) and isentropic (vertical shade) thermal instability
regimes in terms of temperature dependence $\beta$ and density
dependence $\delta$ in the parameterized cooling function. The
squares $\Box$ represent the values of $\beta$ and $\delta$ from
table~1. \label{fig4}}
\end{figure}

Equation (\ref{drift3}) shows that the drift velocity is
inversely proportional to the density so that we can choose, in a
general form, $v_d = \mathrm{const.} \times \rho^{-b}$ where
$b>0$. For a self-gravitating slab, the value of $b$ may be
approximated roughly between $0.5$ to $2.0$. This range of values
should be compared to equation (10) of Nejad-Asghar \&
Ghanbari~(2006), where $b=1+\nu$ was used. The heating due to
ion-neutral friction is spatially dependent (because it depends
on gas density and gas density varies with position). Here we do
not consider this issue in the instability analysis because we
are interested considering the AU-scale condensations that may be
formed from waves with small wavelengths (respect to the cloud
size). In this formalism, equation (37) of SLH for isobaric
thermal instability may be generalized to
\begin{equation}\label{slh37}
  \lambda_{c1}=\frac{\lambda_0}{\sqrt{\frac{\delta+(2b-\nu)\xi}{\beta}-1}}
\end{equation}
with $\xi\equiv \Gamma_{AD}/\Lambda$ and $\lambda_0 \equiv 2 \pi
\sqrt{KT/\rho\beta\Lambda}$ where $K$ is the thermal conduction
coefficient. If the wavelength of perturbation is larger than the
critical wavelength (\ref{slh37}), there will occur isobaric
instability. This requires
\begin{equation}\label{slh46}
  \delta+(2b-\nu)\xi > \beta.
\end{equation}
Since $0<\xi<1$, we find that
\begin{equation}\label{slh48}
  \delta > \beta - (2b-\nu)
\end{equation}
is required for occurrence of isobaric thermal instability. On
the other hand, critical wavelength of isentropic thermal
instability (equation 38 of SLH) is modified as
\begin{equation}\label{slh38}
  \lambda_{c2}=\frac{\lambda_0}{\sqrt{-\frac{3}{2}\frac{\delta-(2b-\nu)\xi}{\beta}-1}},
\end{equation}
which requires
\begin{equation}\label{slh50}
  \delta < -\frac{2}{3} \beta-(2b-\nu)\xi.
\end{equation}
Thus, the condition
\begin{equation}\label{slh51}
  \delta<-\frac{2}{3}\beta
\end{equation}
must at least be held for occurrence of isentropic thermal
instability. The instability regimes are summarized in
Fig.~\ref{fig4}. Depending on the chosen density range, the values
of $\beta$ and $\delta$ can differ, but we can only expect the
isobaric thermal instability mode. The isentropic regime can not
be fulfilled in the molecular slab. Therefore, we only consider
the nonlinear isobaric thermal instability mode.

\begin{figure}
\epsscale{1.0} \plotone{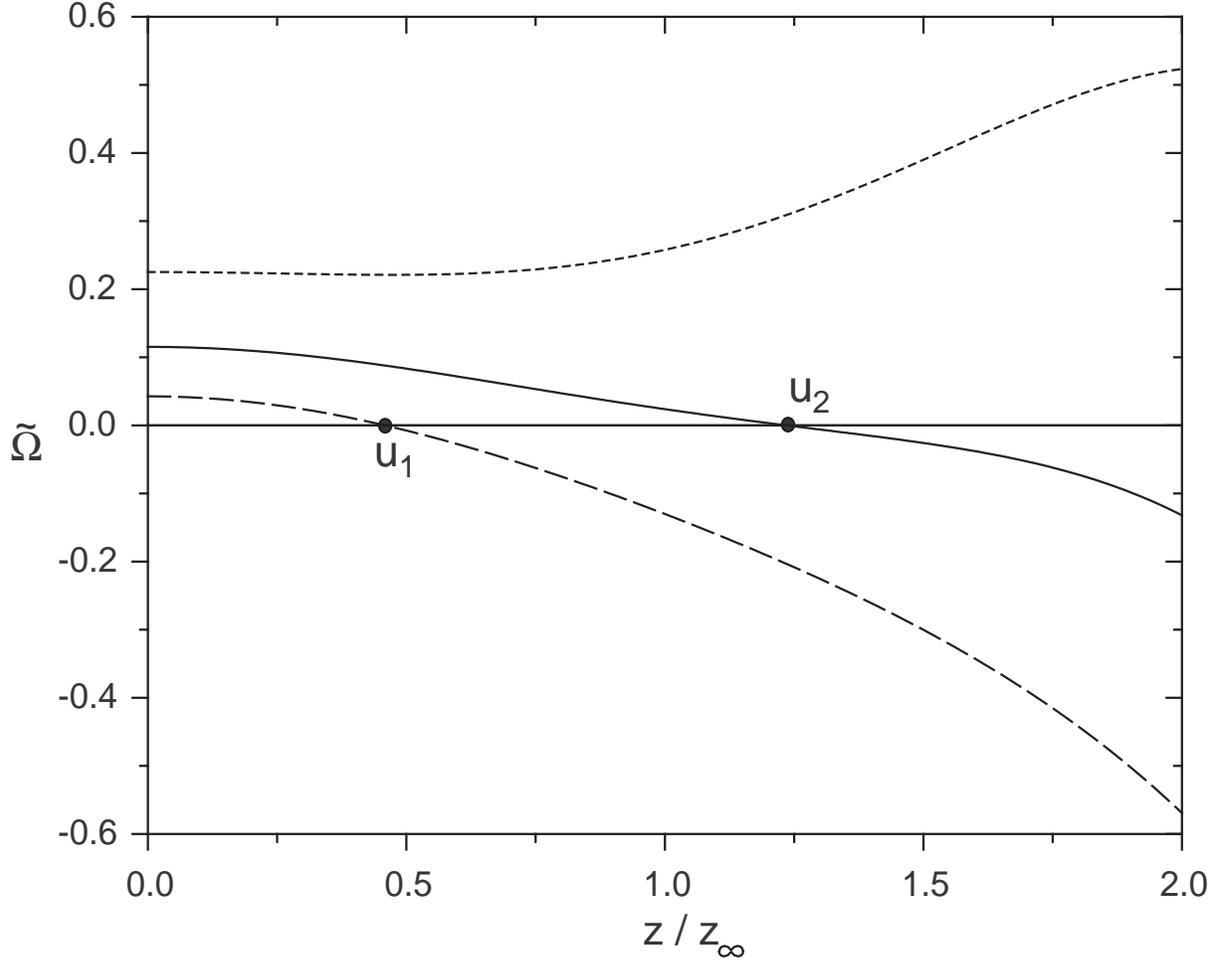} \caption{The non-dimensional net
cooling function of the self-gravitating slab with $\alpha_0=10$,
$a=0.3\mathrm{km.s^{-1}}$, and $n_0=10^{12}\mathrm{m^{-3}}$ for
cases $\tilde{T}_{(\tilde{t}=0)}=4$ (dash), $5$ (solid), and $6$
(dot). The points $u_1$ and $u_2$ are unstable because of
$d\tilde{\Omega}/d\tilde{z}<0$.\label{fig5}}
\end{figure}

\begin{figure}
\epsscale{1.0} \plotone{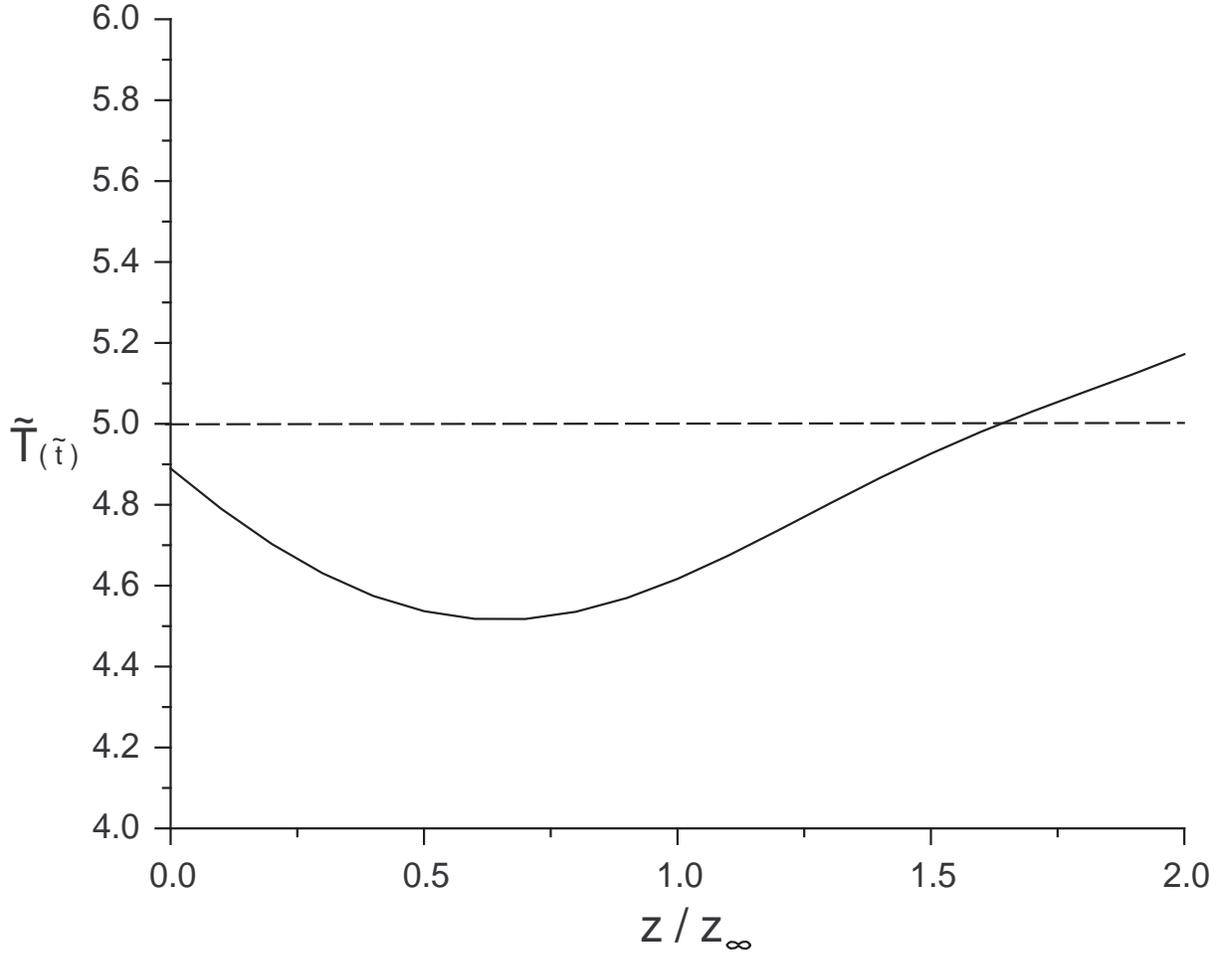} \caption{Evolution of the
temperature of the isobaric unstable regions in self-gravitating
molecular slab with $\alpha_0=10$, $a=0.3\mathrm{km.s^{-1}}$, and
$n_0=10^{12}\mathrm{m^{-3}}$ from initial value $\tilde{T}=5$ at
$\tilde{t}=0$ (dash) to a typical time $\tilde{t}=1$ (solid).
\label{fig6}}
\end{figure}

To give a complete description of the nonlinear thermal runaway,
one needs to solve the nonlinear fluid equations. The energy
equation follows from the use of the first law of thermodynamics,
that is
\begin{equation}\label{energy1}
  \frac{1}{\gamma-1} \frac{dp}{dt} - \frac{\gamma}{\gamma-1}
  \frac{p}{\rho} \frac{d\rho}{dt} + \rho \Omega_{(\rho,T)}=0,
\end{equation}
where $\gamma$ is the polytropic index, $\Omega_{(\rho,T)}$ is
the net cooling function
\begin{equation}\label{netcool}
  \Omega_{(\rho,T)} \equiv \Lambda_{(\rho,T)} -
  (\Gamma_{CR} + \Gamma_{AD}),
\end{equation}
and the pressure $p$ is given by the ideal gas equation of state
\begin{equation}\label{ideal}
  p=\frac{R}{\mu} \rho T,
\end{equation}
where $\mu$ is the mean atomic mass per particle and $R$ is the
universal gas constant. Being interested in the development of
the isobaric thermal instability in the quasi-hydrostatic
molecular slab, we put $dp/dt=0$ in the energy equation
(\ref{energy1}) and obtain the following equation
\begin{equation}\label{energy2}
  \frac{dT}{dt} + \frac{\gamma-1}{\gamma} \frac{\mu}{R} \Omega_{(T,z)}=0,
\end{equation}
where the isobaric condition $\rho \propto T^{-1}$ is used. Now
let us introduce the dimensionless temperature, $\tilde{T}\equiv
T/T_0$, to present the energy equation (\ref{energy2}) as
dimensionless form
\begin{equation}\label{energy3}
  \frac{\partial \tilde{T}}{\partial \tilde{t}} + \tilde{\Omega}_{(\tilde{T},\tilde{z})}=0
\end{equation}
where $\tilde{\Omega}_{(\tilde{T},\tilde{z})}$ is the
dimensionless net cooling function of the slab
\begin{equation}\label{omega0}
  \tilde{\Omega}_{(\tilde{T},\tilde{z})} \equiv (\frac{\gamma-1}{\gamma}
  \frac{\mu}{R}\frac{1}{T_0})
  (\frac{\gamma_{AD} \epsilon}{2(2\pi G)^{1-\nu}}) (\frac{a^{2-2\nu}}{2\pi
  G\sigma_\infty^{2-2\nu}}) \Omega_{(T,z)}.
\end{equation}
Thus, the nonlinear dynamics of the isobaric thermal instability
is determined solely by the form of the net cooling function
curve $\tilde{\Omega}_{(\tilde{T},\tilde{z})}$ and the initial
conditions in the equation (\ref{energy3}). We use the equations
(\ref{heatAD2}), (\ref{heatCR}), and (\ref{cool})-(\ref{beta}) to
find the net cooling function of a self-gravitating molecular
slab. Figure~\ref{fig5} shows the net cooling function of the slab
for typical values of $\tilde{T}_{(\tilde{t}=0)}$. The necessary
condition for the instability has the form $d\tilde{\Omega} /
d\tilde{n}>0$, which in the slab can be expressed as
$d\tilde{\Omega} / d\tilde{z}<0$. This condition can be easily
seen to be fulfilled with the equilibrium points $u_1$ and $u_2$
in the Fig.~\ref{fig5}. Indeed, these results accurately coincide
with the linear instability regimes that are summarized in
Fig.~\ref{fig3}. We consider a constant initial form of the
temperature profile in the slab, and solve numerically the
temperature equation (\ref{energy3}) with approximation that the
dynamics of the slab is solely determined by the equations
(\ref{magcon4})-(\ref{mascon4}) and the initial condition
(\ref{dens}) as outlined by Shu~(1983). Figure~\ref{fig6} gives a
plot of the evolution of the temperature profile in a typical case
described in Fig.~\ref{fig5}. As can be seen in Fig.~\ref{fig6},
large temperature gradients are created with time at the
interfaces between different unstable thermal phases.

\section{Summary and Conclusion}
In this paper the importance of ambipolar drift heating rate on
net cooling function have been investigated, and the isobaric
thermal instability in a self-gravitating magnetized molecular
slab is analyzed. We re-formulate the equations of a
quasi-magnetohydrostatic self-gravitation slab so that the
density dependence of the drift velocity is estimated. Since the
drift velocity is inversely proportional to the density, the
ambipolar drift heating rate is predominated in the outer regions
of the slab.

Choosing a parametric density dependence of drift velocity, the
isentropic and isobaric thermal instability criteria of SLH are
generalized. We have applied the cooling rate of G01 to find the
physical contents of the instability criteria. We have found that
the isentropic thermal instability can not occur in the molecular
cloud, while the isobaric criteria is accurately complied. This
term is also investigated by considering of thermal equilibrium
in pressure-density plane, and by plotting of the net cooling
function in Fig.~\ref{fig5}.

We have considered a constant initial form of the temperature
profile in the slab, and have solved numerically the nonlinear
isobaric energy equation with approximation that the dynamics of
the slab is solely determined by the isothermal equations as
outlined by Shu~(1983). In this way, a large temperature gradient
are created with time at the interfaces between different
isobaric unstable thermal phases. In the context of our thermal
instability analysis in the molecular slab, we find that the
isobaric thermal instability can occur in some regions of it,
therefore it may produce the slab fragmentation and formation of
the spherical AU-scale condensations.

\acknowledgments

\section*{ACKNOWLEDGMENTS}

I appreciate the careful reading and suggested improvements by
Patrick Hennebelle, the reviewer. This work has been supported by
Research Institute for Astronomy and Astrophysics of Maragha
(RIAAM).


\clearpage

\clearpage

\clearpage

\clearpage

\clearpage

\clearpage

\clearpage

\begin{table}
\begin{center}
\caption{Parameters\tablenotemark{a} for the gas cooling function
with standard abundances\tablenotemark{b} and velocity gradient
of $1~\mathrm{km.s}^{-1}.\mathrm{pc}^{-1}$.\label{tbl}}
\begin{tabular}{ccccc}
\tableline\tableline $\log(\frac{n}{n_0})$\tablenotemark{c} &
$\log(\frac{\Lambda_{(n)}}{\mathrm{J.kg^{-1}.s^{-1}}})$ & $\beta_{(n)}$ &
$\log(\frac{\Lambda_0}{\mathrm{J.kg^{-1}.s^{-1}}})$\tablenotemark{d} & $\delta$ \\
\tableline
-4   &-7.82   &1.4 &                       &      \\
     &        &    &$\}\Rightarrow$ -6.14  &0.42  \\
-3.5 &-7.61   &1.8 &                       &      \\
     &        &    &$\}\Rightarrow$ -7.40  &0.06  \\
-3   &-7.58   &2.4 &                       &      \\
     &        &    &$\}\Rightarrow$ -8.45  &-0.29 \\
-2   &-7.87   &2.7 &                       &      \\
     &        &    &$\}\Rightarrow$ -8.65  &-0.39 \\
-1   &-8.26   &3.0 &                       &      \\
     &        &    &$\}\Rightarrow$ -8.96  &-0.7  \\
0    &-8.96   &3.4 &                       &      \\
\tableline
\end{tabular}
\tablenotetext{a}{$\Lambda_{(n,T)}=\Lambda_{(n)}(T/10\mathrm{K})^{\beta_{(n)}}$.}
\tablenotetext{b}{Goldsmith~(2001).}
\tablenotetext{c}{$n_0=10^{12}\mathrm{m^{-3}}$.}
\tablenotetext{d}{$\Lambda_{(n)}=\Lambda_0 (n/n_0)^\delta$.}
\end{center}
\end{table}


\begin{thebibliography}{}
\bibitem[]{aud05} Audit, E., Hennebelle, P., 2005, A\&A, 433, 1
\bibitem[]{bir00} Birk, G.T., 2000, Phys. Plasma, 7, 3811
\bibitem[]{bla87} Black, J.H., 1987, in Interstellar Processes,
                    ed. Hollenbach, D.J., Thronsen, H.A., D.~Reidel
                    Publishing Company, p.~731
\bibitem[]{boi05} Boiss\'{e}, P. Le Petit, F., Rollinde, E., Roueff, E.,
                    Pineau des For\^{e}ts, G., Andersson, B.G., Gry, C., Felenbok,
                    P., 2005, A\&A, 429, 509
\bibitem[]{dic90} Dickman, R.L., Horvath, M.A., Margulis, M.,
                    1990, ApJ, 365, 586
\bibitem[]{fal04} Falgarone, E., Hily-Blant, P., Levrier, F.,
                    2004, Ap\&SS, 292, 89
\bibitem[]{fal02} Falle, S.A.E.G., Hartquist, T.W., 2002, MNRAS,
                    329, 195
\bibitem[]{fol04} Folini, D., Heyvaerts, J., Walder, R., 2004,
                    A\&A, 414, 559
\bibitem[]{gam96} Gammie, C.F., Ostriker, E.C., 1996, ApJ, 466,
                    814
\bibitem[]{gil84} Gilden, D.L., 1984, ApJ, 283, 679
\bibitem[]{gol01} Goldsmith, P.F., 2001, ApJ, 557, 736 (G01)
\bibitem[]{gol78} Goldsmith, P.F., Langer, W.D., 1978, ApJ, 222, 881
\bibitem[]{hei06} Heitsch, F., Slyz, A.D., Devriendt, J.E.G., Hartmann, L.W.,
                    Burkert, A., 2006, ApJ, 648, 1052
\bibitem[]{hen06} Hennebelle, P., Passot, T., 2006, A\&A, 448,
                    1083
\bibitem[]{hen00} Hennebelle, P., Per\'{a}ult, M., 2000, A\&A,
                    359, 1124
\bibitem[]{ino07} Inoue, T., Inutsuka, S., Koyama, H., 2007, ApJ,
                    658, 99
\bibitem[]{koy02} Koyama, H., Inutsuka, S., 2002, ApJ, 564, 97
\bibitem[]{kwa79} Kwan, J., 1979, ApJ, 229, 567
\bibitem[]{kwa86} Kwan, J., Sanders, D.B., 1986, ApJ, 309, 783
\bibitem[]{lis00} Liszt, H., Lucas, R., 2000, A\&A, 355, 333
\bibitem[]{loe90} Loewenstein, M., 1990, ApJ, 349, 471
\bibitem[]{mar93} Marscher, A.P., Moore, E.M., Bania, T.M., 1993, ApJ, 419, 101
\bibitem[]{moo95} Moore, E.M., Marscher, A.P., 1995, ApJ, 452, 679
\bibitem[]{mur96} Murray, S.D., Lin, D.N.C., 1996, ApJ, 467, 728
\bibitem[]{nej06} Nejad-Asghar, M., Ghanbari, J.,  2003, 2003, MNRAS, 345, 1323
\bibitem[]{nej07} Nejad-Asghar, M., Ghanbari, J.,  2006, Ap\&SS, 302, 243
\bibitem[]{neu95} Neufeld, D.A., Leep, S., Melnick, G.J., 1995,
                     ApJS, 100, 132
\bibitem[]{pan01} Pan, K., Federman, S.R., Welty, D.E., 2001, ApJ, 558, 105
\bibitem[]{per85} Perault, M., Falgarone, E., Puget, J.L., 1985,
                     A\&A, 152, 371
\bibitem[]{pri01} Pringle, J.E., Allen, R.J., Lubow, S.H., 2001,
                     MNRAS, 327, 663
\bibitem[]{rol03} Rollinde, E., Boiss\'{e}, P., Federman, S.R., Pan, K., 2003, A\&A, 401, 215
\bibitem[]{shu83} Shu, F., 1983, ApJ, 273, 202
\bibitem[]{shu92} Shu, F., 1992, The Physics of Astrophysics, University Science Books, Mill Valley, CA., Vol~II., p.~360
\bibitem[]{sil79} Silk, J., Takahashi, T., 1979, ApJ, 229, 242
\bibitem[]{ste99} Steele, C.D.,C., Ib\'{a}\~{n}ez S., M.H., 1999, Phys. Plasma,
                     6, 3086
\bibitem[]{sti06} Stiele, H., Lesch, H., Heitsch, F., 2006, MNRAS, 372,
                    862 (SLH)
\bibitem[]{tie97} Tielens, A.G.G.M., Whittet, D.C.B., 1997, IAU
                    Symp. 178, Molecules in Astrophysics: Probes
                    and Processes, ed. van Dishoeck, E.F.,
                    Dordrecht: Kluwer, 45
\bibitem[]{tho96} Thoraval, S., Boiss\'{e}, P., Stark, R., 1996, A\&A, 973, 312
\bibitem[]{vaz06} Vazquez-Semadeni, E., Ryu, D., Passot, T., Gonz\'{a}lez, R.F.,
                    Gazol, A., 2006, ApJ, 643, 245
\bibitem[]{wil00} Williams, J.P., Blitz, L., McKee, C.F., 2000, in Protostars and Planets
                     IV, ed. Mannings, V., Boss, A.P., Russel, S.S.,
                     Tucson: University of Arizona Press, p.~97
\end{thebibliography}
\end{document}